\documentclass[conference, final]{IEEEtran}
\usepackage[nolist]{acronym} 
\usepackage{tikz}
\usepackage{pgfplots}
\usepgfplotslibrary{groupplots}
\usepackage{graphicx}
\usepackage{subfloat}
\usepackage{subfig}
\usepackage{caption}
\usepackage{float}
\pgfplotsset{compat=newest}
\usepackage{units}
\usepackage{amsmath, amsbsy, amssymb}
\usepackage{tikzscale}
\usetikzlibrary{arrows}
\usepackage{color}
\usepackage{epigraph}
\usepackage{circuitikz}
\usepackage{multirow}
\usepackage{booktabs}
\usepackage[ruled,boxed,lined]{algorithm2e}
\usepackage[group-separator={,}]{siunitx}
\definecolor{darkgreen}{rgb}{0.125,0.5,0.169}
\usetikzlibrary{shapes,arrows}
\usepackage{marvosym}
\usepackage{bbm}

\tikzset{>=latex}

\renewcommand{\vec}[1]{\mathbf{#1}}

\newcommand{\nv}{\vec{n}}

\newcommand{\pv}{\vec{p}}

\newcommand{\uv}{\vec{u}}

\newcommand{\xv}{\vec{x}}
\newcommand{\yv}{\vec{y}}

\begin{acronym}
 \acro{CSI}{channel state information}
 \acro{UE}{user equipment}
 \acro{UL}{uplink}
 \acro{BS}{basestation}
 \acro{TDD}{time division duplex}
 \acro{FDD}{frequency division duplex}
 \acro{ECC}{error-correcting code}
 \acro{MLD}{maximum likelihood decoding}
 \acro{HDD}{hard decision decoding}
 \acro{IF}{intermediate frequency}
 \acro{RF}{radio frequency}
 \acro{SDD}{soft decision decoding}
 \acro{NND}{neural network decoding}
 \acro{CNN}{convolutional neural network}
 \acro{ML}{maximum likelihood}
 \acro{GPU}{graphical processing unit}
 \acro{BP}{belief propagation}
 \acro{LTE}{Long Term Evolution}
 \acro{BER}{bit error rate}
 \acro{SNR}{signal-to-noise-ratio}
 \acro{ReLU}{rectified linear unit}
 \acro{BPSK}{binary phase shift keying}
 \acro{QPSK}{quadrature phase shift keying}
 \acro{AWGN}{additive white Gaussian noise}
 \acro{MSE}{mean squared error}
 \acro{LLR}{log-likelihood ratio}
 \acro{MAP}{maximum a posteriori}
 \acro{NVE}{normalized validation error}
 \acro{BCE}{binary cross-entropy}
 \acro{BLER}{block error rate}
 \acro{SQR}{signal-to-quantisation-noise-ratio}
 \acro{MIMO}{multiple-input multiple-output}
 \acro{OFDM}{orthogonal frequency division multiplex}
 \acro{RF}{radio frequency}
 \acro{LOS}{line of sight}
 \acro{NLoS}{non-line of sight}
 \acro{NMSE}{normalized mean squared error}
 \acro{CFO}{carrier frequency offset}
 \acro{SFO}{sampling frequency offset}
 \acro{IPS}{indoor positioning system}
 \acro{TRIPS}{time-reversal IPS}
 \acro{RSSI}{received signal strength indicator}
 \acro{MIMO}{multiple-input multiple-output}
 \acro{ENoB}{effective number of bits}
 \acro{AGC}{automated gain control}
 \acro{ADC}{analog to digital converter}
 \acro{ADCs}{analog to digital converters}
 \acro{FB}{front bandpass}
 \acro{FPGA}{field programmable gate array}
 \acro{JSDM}{Joint Spatial Division and Multiplexing}
 \acro{NN}{neural network}
 \acro{IF}{intermediate frequency}
 \acro{LoS}{line-of-sight}
 \acro{NLoS}{non-line-of-sight}
 \acro{DSP}{digital signal processing}
 \acro{AFE}{analog front end}
 \acro{SQNR}{signal-to-quantisation-noise-ratio}
 \acro{SINR}{signal-to-interference-noise-ratio}
 \acro{ENoB}{effective number of bits}
 \acro{AGC}{automated gain control}
 \acro{PCB}{printed circuit board}
 \acro{EVM}{error vector mangnitude}
 \acro{CDF}{cumulative distribution function}
 \acro{MRC}{maximum ratio combining}
 \acro{MRP}{maximum ratio precoding}
 \acro{MRT}{maximum ratio transmission}
 \acro{DeepL}{deep-learning}
 \acro{DL}{deep learning}
 \acro{SISO}{single-input single-output}
 \acro{SGD}{stochastic gradient descent}
 \acro{CP}{cyclic prefix}
 \acro{MISO}{Multiple Input Single Output}
 \acro{LMMSE}{linear minimum mean square error}
 \acro{ZF}{zero forcing}
 \acro{USRP}{universal software radio peripheral}
 \acro{RNN}{recurrent neural network}
 \acro{GRU}{gated recurrent unit}
 \acro{LSTM}{long short-term memory}
 \acro{NTM}{neural turing machine}
 \acro{DNC}{differentiable neural computer}
 \acro{TCN}{temporal convolutional network}
 \acro{FCL}{fully connected layer}
 \acro{MANN}{memory augmented neural network}
 \acro{RNN}{recurrent neural network}
 \acro{DNN}{dense neural network}
 \acro{FIR}{finite impulse response}
 \acro{BPTT}{back-propagation through time}
 \acro{GAN}{generative adversarial network}
 \acro{ELU}{exponential linear unit}
 \acro{tanh}{hyperbolic tangent}
 \acro{BICM}{bit-interleaved coded modulation}
\end{acronym}

\definecolor{mittelblau}{RGB}{0, 126, 198}
\definecolor{violettblau}{cmyk}{0.9, 0.6, 0, 0}
\definecolor{rot}{RGB}{238, 28 35}
\definecolor{apfelgruen}{RGB}{140, 198, 62}
\definecolor{gelb}{RGB}{1, 221, 0}
\definecolor{orange}{RGB}{244, 111, 33}
\definecolor{pink}{RGB}{237, 0, 140}
\definecolor{lila}{RGB}{128, 10, 145}
\definecolor{hellgrau}{RGB}{224, 224, 224}
\definecolor{mittelgrau}{RGB}{128, 128, 128}
\definecolor{dunkelgrau}{RGB}{80,80,80}
\definecolor{anthrazit}{RGB}{19, 31, 31}

\pgfplotscreateplotcyclelist{corporate colours markers}{%
rot, every mark/.append style={fill=.!80!rot},mark=*\\%
mittelblau, every mark/.append style={fill=.!80!mittelblau},mark=square*\\%
apfelgruen, every mark/.append style={fill=.!80!apfelgruen},mark=triangle*\\%
orange, mark=star\\%
pink, every mark/.append style={fill=.!80!pink},mark=diamond*\\%
violettblau, every mark/.append style={fill=.!80!violettblau},mark=otimes*\\%
lila, mark=|\\%
gelb, every mark/.append style={fill=.!80!gelb},mark=pentagon*\\%
hellgrau, mark=text,text mark=p\\%
anthrazit, mark=text,text mark=a\\%
}

\pgfplotscreateplotcyclelist{corporate colours markers double}{%
rot, every mark/.append style={fill=.!80!rot},mark=*\\%
rot, dashed, every mark/.append style={fill=.!80!rot,solid},mark=*\\%
mittelblau, every mark/.append style={fill=.!80!mittelblau},mark=square*\\%
mittelblau, dashed, every mark/.append style={fill=.!80!mittelblau,solid},mark=square*\\%
apfelgruen, every mark/.append style={fill=.!80!apfelgruen},mark=triangle*\\%
apfelgruen, dashed, every mark/.append style={fill=.!80!apfelgruen,solid},mark=triangle*\\%
orange, mark=star\\%
orange, dashed, mark=star\\%
pink, every mark/.append style={fill=.!80!pink},mark=diamond*\\%
pink, dashed, every mark/.append style={fill=.!80!pink,solid},mark=diamond*\\%
violettblau, every mark/.append style={fill=.!80!violettblau},mark=otimes*\\%
violettblau, dashed, every mark/.append style={fill=.!80!violettblau,solid},mark=otimes*\\%
lila, mark=|\\%
lila, dashed, mark=|\\%
gelb, every mark/.append style={fill=.!80!gelb},mark=pentagon*\\%
gelb, dashed, every mark/.append style={fill=.!80!gelb,solid},mark=pentagon*\\%
}

\newcommand\deactreview[1]{#1}%

\IEEEoverridecommandlockouts

\begin{document}

\title{On Recurrent Neural Networks for Sequence-based Processing in Communications}

\author{\IEEEauthorblockN{Daniel Tandler, Sebastian D\"orner, Sebastian Cammerer, and Stephan ten Brink\\}%
\IEEEauthorblockA{
Institute of Telecommunications, Pfaffenwaldring 47, University of  Stuttgart, 70659 Stuttgart, Germany \\ \{doerner,cammerer,tenbrink\}@inue.uni-stuttgart.de
}
\deactreview{\thanks{This work has been supported by DFG, Germany, under grant BR 3205/6-1. 
We also acknowledge support from NVIDIA through their academic program which granted us a TITAN V graphics card that was used for this work.}}
}

\maketitle

\begin{abstract}

In this work, we analyze the capabilities and practical limitations of \acp{NN} for sequence-based signal processing which can be seen as an omnipresent property in almost any modern communication systems.
In particular, we train multiple state-of-the-art \ac{RNN} structures to \emph{learn how to decode} convolutional codes allowing a clear benchmarking with the corresponding \ac{ML} Viterbi decoder.
We examine the decoding performance for various kinds of \ac{NN} architectures, beginning with \emph{classical} types like feedforward layers and \ac{GRU}-layers, up to more recently introduced architectures such as \acp{TCN} and \acp{DNC} with external memory.
As a key limitation, it turns out that the training complexity increases exponentially with the length of the encoding memory $\nu$ and, thus, practically limits the achievable \ac{BER} performance. 
To overcome this limitation, we introduce a new training-method by gradually increasing the number of \emph{ones} within the training sequences, i.e., we constrain the amount of possible training sequences in the beginning until first convergence.
By consecutively adding more and more possible sequences to the training set, we finally achieve training success in cases that did not converge before via \emph{naive} training.
Further, we show that our network can learn to jointly detect and decode a \ac{QPSK} modulated code with sub-optimal (anti-Gray) labeling in \emph{one-shot} at a performance that would require iterations between demapper and decoder in classic detection schemes.

\end{abstract}

\deactreview{\acresetall} %
\section{Introduction}

The huge success of \ac{DL} and \acp{NN}, mainly in the fields of computer vision and speech processing, has recently triggered further exploration of \ac{DL} for communications.
The potential applications span from trainable channel decoders \cite{nachmani2016learning,yihan2019deepturbo,kim2018communication,jiang2019mind,gruber2017deep}, \ac{NN}-based \ac{MIMO} detectors \cite{samuel2017deep} and detectors for molecular channels \cite{farsad2017detection} up to communication systems that inherently \emph{learn to communicate} \cite{o2017introduction}.
Most of these applications typically rely on block-based signal processing, which is an obvious consequence if \ac{NN} structures are used that were derived from computer vision tasks.
Besides, those structures also benefit from the possibility of straightforward \ac{SGD}-based training.

In the contrary to that, signal processing for communications often benefits from sequence-based processing (e.g., a simple \ac{FIR} filter for equalization) as it allows to maintain an internal state.
Mainly driven by the speech processing community, a rich variety of different sequence-based \ac{RNN} structures emerged, which can typically also be trained by \ac{SGD} when truncated \ac{BPTT} is used.
In \cite{farsad2018neural}, advantages for detection over molecular channels have been reported and the authors of \cite{kim2018communication,lyu2018performance} show that \acp{RNN} can improve the performance of channel decoding and also code design \cite{jiang2018learn}.

In this work, we aim to compare different families of \ac{RNN} architectures with fundamentally different properties (e.g., \acp{NN} with external memory) rather than minor implementation differences (e.g., \ac{GRU} vs. \ac{LSTM}).
This is also supported by \cite{greff2017lstm} as the authors have shown by exhaustive search that the average performance of the \ac{GRU} or \ac{LSTM} cell structure does not significantly differ \cite{greff2017lstm}.

\deactreview{On the other hand, convolutional codes are well-understood since many years and can be seen as the workhorse of many communication systems \cite{forney1974convolutional,hemmati1977truncation,ieee_wifi_n,3gpp_LTE_advanced}.}
Besides their simple encoding structures, convolutional codes benefit from the availability of an \ac{ML} decoder, namely the well-known Viterbi algorithm \cite{viterbi1967error}.
Thus, convolutional codes allow an easy benchmark by providing a clear (and optimal) baseline to analyze the influence of encoding memory and traceback-length, i.e., how close can a given \ac{NN} approximate the optimal decoder for specifically chosen constraints.

The universal approximator theorem \cite{hornik1989multilayer}, and the fact that an explicit optimal algorithm exists, directly tells us that also an \ac{NN} must exist (neglecting any complexity constraints) which comes arbitrarily close to the optimal performance.
While other groups already showed the existence of such decoding \acp{NN} for short memory convolutional codes \cite{yihan2019deepturbo,kim2018communication}, we want to further investigate to what extent \acp{NN} are capable of processing even more complex information sequences using the example of decoding convolutional codes up to memory $\nu$.
However, in practice the limiting factor is clearly the training complexity and, thus, the major challenge is to find a suitable training method for this task.
This is also why we have made parts of the source code of this paper available\footnote{Source code available at: https://github.com/sdnr/RNN-Conv-Decoder}, as we hope it could be useful for others working in communications at their specific processing tasks.
We want to point out that the aim of this work is \emph{not} to outperform the Viterbi decoder, but to provide insights into a suitable training methodology and efficient \ac{NN} architectures for continuous signal processing in communications.
Yet, a potential benefit can be seen in other metrics like the possibility of \emph{learning to approximate} a (sub-optimal but) low-complex decoder for prohibitively large encoding memories (cf. the \emph{NASA Big Viterbi Decoder} \cite{onyszchuk1991coding}).

In \cite{gruber2017deep}, it has been shown for block-codes, that \acp{NN} are limited by an exponential training complexity when training with all possible codewords is required, i.e., for $k$ information bits $2^k$ different codewords needed to be shown during training.
In the case of convolutional codes there is a naturally limited length that still allows (close to) optimal decoding (cf. \emph{traceback length} in Viterbi decoding \cite{forney1974convolutional,hemmati1977truncation}).

Moreover, we believe both \ac{NN} structures and the training procedure, as shown for convolutional codes in this paper, are of significant practical importance in many other deep-learning-based communication applications like equalization, continuous \ac{CSI} prediction and also with regard to the scalability of autoencoder-driven systems \cite{o2017introduction}.

\section{System Model and Neural Network Architectures}

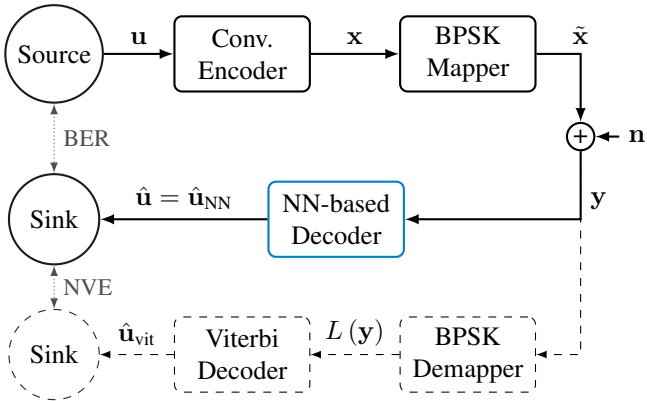
\begin{figure}
\centering
\begin{tikzpicture}
\tikzstyle{box} = [draw,rounded corners=.1cm,minimum height=1cm,minimum width=1.8cm, align=center] 
\tikzstyle{circ} = [draw=anthrazit,circle,minimum size=1.2cm, text centered,align=center] 

\node[circ,thick] (src) at (0,0){Source};
\node[box,thick] (cenc) at (2.5,0) {Conv.\\Encoder};
\node[box,thick] (mapper) at (5.5,0) {BPSK\\Mapper};
\node[circle,thick,draw,inner sep=1] (ch) at (7,-1.1) {+};
\node (noise) at (7.75,-1.1) {$\nv$};
\node[box,thick,draw=mittelblau] (cdec) at (3.75,-2.2) {NN-based\\Decoder};
\node[circ,thick] (sink) at (0,-2.2){Sink};
\node[dashed,box] (vit) at (2.5,-4.0) {Viterbi\\Decoder};
\node[dashed,box] (demap) at (5.5,-4.0) {BPSK\\Demapper};
\node[dashed,circ] (vit_sink) at (0,-4.0){Sink};

\draw [->,thick]  (src.east) --  (cenc.west) node[midway,anchor=south]{$\uv$};
\draw [->,thick]  (cenc.east) -- (mapper.west) node[midway,anchor=south]{$\xv$};
\draw [->,thick]  (mapper.east) -| (ch.north) node[midway,anchor=south]{$\tilde{\xv}$};
\draw [->,thick]  (noise.west) -- (ch.east);
\draw [->,thick]  (ch.south) |- (cdec.east) node[midway,anchor=south west]{$\yv$};
\draw [->,thick]  (cdec.west) -- (sink.east) node[midway,anchor=south]{$\hat{\uv}=\hat{\uv}_\text{NN}$};
\draw [->,dashed]  (ch.south) |- (demap.east);
\draw [->,dashed]  (demap.west) -- (vit.east) node[midway,anchor=south]{$L \left( \yv \right)$};
\draw [->,dashed]  (vit.west) -- (vit_sink.east) node[midway,anchor=south]{$\hat{\uv}_\text{vit}$};

\draw [<->,densely dotted,dunkelgrau] (src.south) -- (sink.north) node[midway,anchor=west]{\small BER};
\draw [<->,densely dotted,dunkelgrau] (sink.south) -- (vit_sink.north) node[midway,anchor=west]{\small NVE};

\end{tikzpicture}
\caption{System model}
\label{fig:system_model}
\vspace*{-0.3cm}
\end{figure}

Due to many breakthroughs and rapidly increasing research in machine learning for various domains, many different network structures and concepts have emerged.
\deactreview{Each of those domains developed its own key \ac{NN} layer architecture to successfully cope with the specific tasks.
Most famously, \emph{convolutional} layers revolutionized the field of computer vision, \emph{recurrent} layers, in combination with \ac{BPTT}, enhanced natural language and speech processing, and currently \acp{GAN} are even aspiring to take one of the last human bastions -- that is -- creativity.}
When taking a closer look at the characteristics of most signals in the domain of communications we find:
\begin{itemize}
\item Sequences: signals are sequential, but can be often processed in a block-wise manner (cf. traceback).
\item Locality: Single samples are heavily entangled in time, but, unlike sentences or audio speech (where complex context connections over long time distances occur) these dependencies -- with limited memory -- are short and often constant in time (tapped delay, multidimensional modulation, sampling effects).
\item Complex-valued: Their dimensionality is either a single or multiple parallel complex-valued streams of samples that can be represented by concatenating real and imaginary values.
\end{itemize}
Therefore, by seeking the optimal \ac{NN} structure, we focus on recurrent sequence-to-sequence models. 
As we chose to exemplarily decode convolutional codes, our input signals for the decoding \ac{NN} are of the following properties:
\begin{itemize}
\item Two received samples represent one uncoded bit, as we consistently use rate $r = \nicefrac{1}{2}$ codes throughout this work.
\item The original information of the uncoded bit is diffused over several received samples depending on the memory $\nu$ of the applied convolutional code. As a rule of thumb, the affected sequence is of length ``traceback'' \cite{forney1974convolutional,hemmati1977truncation} $\ell_{\text{tb}} \approx 5 \cdot (\nu+1).$
\end{itemize}

Fig.~\ref{fig:system_model} depicts our basic system model where a convolutional encoder maps a stream $\uv$ of uncoded bit $u_k \in \{0,1\}$ to a stream $\xv$ of coded bit $x_k \in \{0,1\}$, a mapper that maps those coded bits to a stream $\tilde{\xv}$ of \ac{BPSK} symbols $\tilde{x}_k \in \{-1,1\}$, an \ac{AWGN} channel with output $\yv = \tilde{\xv} + \nv$, where $\nv \sim \mathcal{N}(0,\sigma^{2})$, and finally the \ac{NN} decoder that predicts $\hat{\uv}$, defined as $\hat{u}_k \in \{0,1\}$, given $\yv$ by inherently adopting a demapping scheme.
It also shows the Viterbi baseline system in dashed lines, where $L \left( \yv \right)$ are the \ac{LLR} values of $\yv$ and $\hat{\uv}_\text{vit}$ is the estimate of the Viterbi decoder.

\subsection{Basic Neural Network Decoder Architecture}
Based on reported experiences \cite{kim2018communication,farsad2018neural,jiang2018learn,lyu2018performance} while facing similar processing problems, and also confirmed by our own empirical experiments, we finally arrived at a baseline decoder layer architecture that makes use of bidirectional state propagation, i.e, processing the sequence from both sides.

\begin{figure}
\centering
\begin{tikzpicture}
\tikzstyle{box} = [draw,rounded corners=.1cm, align=center] 
\tikzstyle{circ} = [draw=anthrazit,circle,minimum size=1.2cm, text centered,align=center] 
\tikzstyle{bits} = [draw=anthrazit, text centered,align=center]

\def\width{3.625};

\draw[hellgrau] (-\width-0.25,10) -- (-\width,10);
\draw[hellgrau] (-\width-0.25,9.75) -- (-\width,9.75);
\draw[hellgrau] (-\width-0.25,9.5) -- (-\width,9.5);

\draw[hellgrau!50] (-\width-0.5,10) -- (-\width-0.25,10);
\draw[hellgrau!50] (-\width-0.5,9.75) -- (-\width-0.25,9.75);
\draw[hellgrau!50] (-\width-0.5,9.5) -- (-\width-0.25,9.5);

\draw[hellgrau!25] (-\width-0.75,10) -- (-\width-0.5,10);
\draw[hellgrau!25] (-\width-0.75,9.75) -- (-\width-0.5,9.75);
\draw[hellgrau!25] (-\width-0.75,9.5) -- (-\width-0.5,9.5);

\draw[hellgrau] (\width+0.25,10) -- (\width,10);
\draw[hellgrau] (\width+0.25,9.75) -- (\width,9.75);
\draw[hellgrau] (\width+0.25,9.5) -- (\width,9.5);

\draw[hellgrau!50] (\width+0.5,10) -- (\width+0.25,10);
\draw[hellgrau!50] (\width+0.5,9.75) -- (\width+0.25,9.75);
\draw[hellgrau!50] (\width+0.5,9.5) -- (\width+0.25,9.5);

\draw[hellgrau!25] (\width+0.75,10) -- (\width+0.5,10);
\draw[hellgrau!25] (\width+0.75,9.75) -- (\width+0.5,9.75);
\draw[hellgrau!25] (\width+0.75,9.5) -- (\width+0.5,9.5);

\draw[hellgrau!50] (-\width-0.5,10) -- (-\width-0.5,9.5);
\draw[hellgrau!25] (-\width-0.75,10) -- (-\width-0.75,9.5);
\draw[hellgrau!50] (\width+0.5,10) -- (\width+0.5,9.5);
\draw[hellgrau!25] (\width+0.75,10) -- (\width+0.75,9.5);

\node[rotate=90] (test) at (-\width-0.015,10.02) {\Cutleft};
\node[rotate=-90] (test) at (\width+0.02,10.02) {\Cutright};

\foreach \x in {0,...,31}
	\draw[hellgrau] (\x*0.25-\width-0.25,10) -- (\x*0.25-\width-0.25,9.5);
\draw[hellgrau] (-\width,9.75) -- (\width,9.75);
\draw (-\width,10) rectangle (\width,9.5);
\node (y) at (0,9.75) {$\yv$};
\node[mittelgrau] (y) at (-1.6,9.75) {\emph{sequence snippet}};

\foreach \x in {0,...,14}
	\draw[->,mittelgrau] (\x*0.5-\width+0.125,9.5) -- (\x*0.5-\width+0.125,9);

\draw[lila,very thick,rounded corners=.1cm,fill=lila!03] (-\width,9) rectangle (\width,8);
\node[anthrazit] (y) at (0,8.5) {Multi-RNN Cell};
\draw[->,rot!40,line width=2] (-\width+0.05,8.75) -- (\width-0.05,8.75);
\draw[<-,mittelblau!40,line width=2] (-\width+0.05,8.25) -- (\width-0.05,8.25);

\foreach \x in {0,...,14}
	\draw[->,mittelgrau] (\x*0.5-\width+0.125,8) -- (\x*0.5-\width+0.125,7.5);

\draw[white,fill=rot!05] (-\width,7.5) rectangle (-0.5*\width-0.25-0.0625,6.75);
\draw[white,fill=mittelblau!05] (-\width,6.75) rectangle (-0.5*\width-0.25-0.0625,6.0);
\draw[white,fill=rot!10] (-0.5*\width-0.25-0.0625,7.5) rectangle (0.5*\width+0.25+0.0625,6.75);
\draw[white,fill=mittelblau!10] (-0.5*\width-0.25-0.0625,6.75) rectangle (0.5*\width+0.25+0.0625,6.0);
\draw[white,fill=rot!05] (0.5*\width+0.25+0.0625,7.5) rectangle (\width,6.75);
\draw[white,fill=mittelblau!05] (0.5*\width+0.25+0.0625,6.75) rectangle (\width,6.0);

\foreach \x in {0,...,28}
	\draw[hellgrau] (\x*0.25-\width,7.5) -- (\x*0.25-\width,6);
\draw[hellgrau] (-\width,7.25) -- (\width,7.25);
\draw[hellgrau] (-\width,7.0) -- (\width,7.0);
\draw[hellgrau] (-\width,6.75) -- (\width,6.75);
\draw[hellgrau] (-\width,6.5) -- (\width,6.5);
\draw[hellgrau] (-\width,6.25) -- (\width,6.25);
\draw (-\width,7.5) rectangle (\width,6);

\foreach \x in {0,...,7}
	\node[box,rotate=90] (dnn\x) at (\x*0.5-0.5*\width+0.0625,5) {DNN};
\foreach \x in {0,...,7}
	\draw[->,mittelgrau] (\x*0.5-0.5*\width+0.0625,6) -- (dnn\x.east);
\foreach \x in {0,...,7}
	\draw[->,mittelgrau]  (dnn\x.west) -- (\x*0.5-0.5*\width+0.0625,4);

\foreach \x in {0,...,15}
	\draw[hellgrau] (\x*0.25-0.5*\width-0.0625,4) -- (\x*0.25-0.5*\width-0.0625,3.75);
\draw (-0.5*\width-0.25-0.0625,4) rectangle (0.5*\width+0.25+0.0625,3.75);
\node (y) at (-0.5*\width-1.45,3.9) {soft output $\hat{\pv}$};

\draw[thick,dashed,rot] (-0.5*\width-0.25-0.0625,7.5) -- (-0.5*\width-0.25-0.0625,4);
\draw[thick,dashed,rot] (0.5*\width+0.25+0.0625,7.5) -- (0.5*\width+0.25+0.0625,4);

\draw[dunkelgrau,dashed] (-0.5*\width-0.25-0.0625,10.35) -- (-0.5*\width-0.25-0.0625,10);
\draw[dunkelgrau,dashed] (0.5*\width+0.25+0.0625,10.35) -- (0.5*\width+0.25+0.0625,10);

\node at (-2.875,10.25) {\small $\ell_{\text{ramp}}$};
\node at (0,10.25) {\small $\ell_{\text{ld}}$};
\node at (2.875,10.25) {\small $\ell_{\text{ramp}}$};
\node[rotate=-90] at (\width+0.25,9.75) {\small $\nicefrac{1}{r}$};
\node[rotate=-90] at (\width+0.25,6.75) {\small $2 \cdot \ell_{\text{rc}}$};

\node[rotate=90] (test) at (-0.5*\width-0.25-0.075,5.98) {\Cutright};
\node[rotate=-90] (test) at (0.5*\width+0.25+0.08,5.98) {\Cutleft};

\end{tikzpicture}
\vspace*{-0.3cm}
\caption{Neural network based decoder architecture.}
\label{fig:network_architecture}
\vspace*{-0.3cm}
\end{figure}
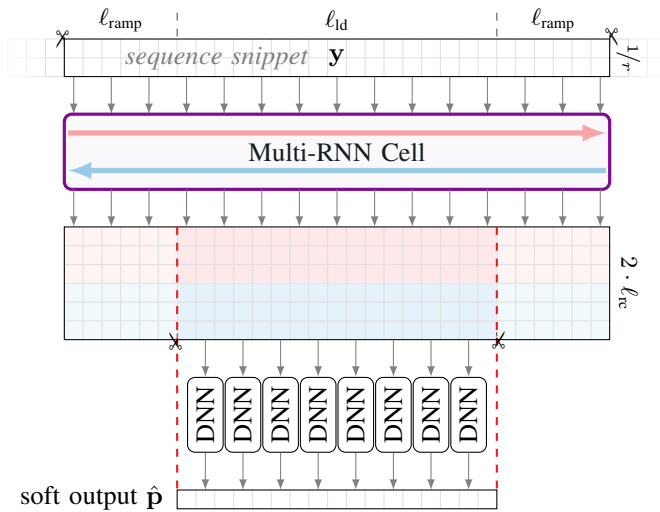

Fig.~\ref{fig:network_architecture} depicts our final \ac{NN}-based decoder architecture which essentially consists of an \ac{RNN} part in the early layers and a \ac{DNN} part in the later layers.
As can be seen, the input to the decoder is a \emph{sequence snippet} $\yv^{\left(2 \ell_{\text{ramp}}+\ell_{\text{ld}}\right) \times 2}$ of $2 \ell_{\text{ramp}}+\ell_{\text{ld}}$ time steps, each containing two noisy channel observations $y$ due to the rate $r = \nicefrac{1}{2}$ code\footnote{We assume $\nicefrac{1}{r}$ is integer, which is typically the case for convolutional codes due to their encoder structure. Otherwise we suggest feeding $\lceil \nicefrac{1}{r} \rceil$ channel observations per time step while either repeating observations or zero padding observations at every second time step.}.
Each time step is sequentially processed in forward and backward direction by a \emph{multi-\ac{RNN} cell}, i.e., a higher level cell structure consisting of several \ac{RNN} layers.
While the output of the multi-\ac{RNN} cell for each time step consists of $\ell_\text{rc}$ values (inspired by \cite{yihan2019deepturbo}), the cell's outputs from both directions are concatenated for each time step. The intuition behind this procedure is to interpret the \ac{RNN} output as a \emph{latent feature} variable which then needs to be further transformed into the decision on bit $\hat{u}_k$.
To mitigate degrading effects during state build up at the beginning and, since we are processing bidirectionally, at the end of the original input sequence, we discard the beginning $\ell_\text{ramp}$ ``ramp-up'' and the last $\ell_\text{ramp}$ ``ramp down'' time step outputs.
Hence, the output tensor of the decoder's recurrent part is of shape $\ell_{\text{ld}} \times 2\ell_\text{rc}$. %

Note that this can be straightforwardly extended to a \emph{stateful} architecture (at least in forward direction) that promises a higher throughput by passing the forward state from sequence-to-sequence instead of discarding it. The intuition behind is to avoid the rebuild of the \emph{internal state} of the decoder similar to the traceback in Viterbi decoding.
We do not follow this approach as it heavily complicates feeding and backward direction processing while not gaining significantly in terms of the final \ac{BER}.
One should also note that both designs based on bidirectional processing, stateless and stateful, exhibit a structural decoding delay of at least $\ell_\text{ramp}$ time steps, caused by the backward processing branch.

The output of the multi-\ac{RNN} cell is then forwarded to a \ac{DNN} layer with $N_\text{DNN}$ units and \ac{ELU} activation function.
This \ac{DNN} layer has no connections through time and is simply meant to combine the ``features'' that were extracted by the preceding multi-\ac{RNN} cell in forward and backward direction.
Finally, the combining layer's output tensor of shape $\ell_{\text{ld}} \times N_\text{DNN}$ is fed into a sigmoid activated layer with only one single neuron to give an estimate $\hat{p}_k = P \left( \hat{u}_k = 1 \right)$, i.e., on the soft-value of $\hat{u}_k$.
Thereby, the final output of our \ac{NN} based decoder is the vector $\hat{\pv}^{\ell_{\text{ld}} \times 1}$  which can be hard decided for \ac{BER} calculations to $\hat{\uv}$, 
where $\hat{u}_k = \mathbbm{1}_{\{\hat{p}_k > 0.5\}}$
and $\mathbbm{1}_{\{x>\delta\}}$ denotes the indicator function, i.e., returns 1 if $x>\delta$ and 0 otherwise.

\subsection{Recurrent Neural Network Cell Structures}

As illustrated by Fig.~\ref{fig:network_architecture}, the core element within our decoder architecture is the recurrent part, namely the multi-\ac{RNN} cell.
Thus, it is of great importance to find a good structure of these \ac{RNN} layers to build up this multi-\ac{RNN} wrapper cell.
Such an \ac{RNN} cell must be capable of generalizing to the task, while still complying to certain complexity constraints to be able to fully train the decoder within a reasonable amount of time.
Out of the broad selection of \ac{NN} structures available, we investigate the most promising ones for our task\deactreview{\footnote{As mentioned in the introduction, we try to cover a wide range of different architectures.}}:

\subsubsection{Fully Connected Dense Layers (\acsp{DNN})\acused{DNN}}

To provide a fair comparison to non-recurrent networks, we also investigate \emph{classical} feed-forward \acp{DNN} without any connections through time.
Thereby, we distinguish between feeding only a single time step to the \ac{DNN} to make a prediction on a single bit (as sanity check), which will not work because the \ac{DNN} has no memory nor recurrent connections through time, and feeding a \emph{snippet} of several time steps to give a prediction on several bits.

\subsubsection{Temporal Convolutional Networks (\acsp{TCN})\acused{TCN} and Trellisnets}

Conventional \acp{CNN} have been mainly used in multidimensional applications, e.g., image classification.
However it has recently been shown \cite{BaiTCN2018}, that they can be successfully applied to one-dimensional sequence-to-sequence task as well, while maintaining their causality by applying a specific amount of padding to the inputs of the convolutional layers.
The use of convolutional layers in \acp{TCN}\footnote{Strictly speaking, \acp{TCN} do not belong to the class of \acp{RNN}, however, we believe it is worth analyzing these sequence-to-sequence models.} leads to parameter-sharing across layers, i.e., the number of parameters is independent of the length of the input sequence.
TrellisNets \cite{BaiTrellis2018} are an extension of \acp{TCN} with the main difference being that the weights are not only shared across single layers, but also between all layers in the network and the input to the network is injected at each layer.

\subsubsection{Memory Augmented Neural Networks (\acsp{MANN})\acused{MANN}}

The class of \acp{MANN} extends the concept of \acp{RNN} with an external memory, with which the network can interact via some sorts of interfaces.
One prominent example of a \ac{MANN} is the \ac{DNC} \cite{graves2016hybrid}. %
\acp{DNC} use a controller, consisting of a traditional feedforward network or recurrent network, to interact with the external memory via read-and write heads. A read-head can read from the memory at each time-step while write-heads can write to the memory at each time step. All operations performed within the \ac{DNC} are fully differentiable, leading \acp{DNC} to be trainable with \ac{BPTT}.
Thus, we use \acp{DNC} with \ac{DNN}- and \ac{GRU}-based controllers, denoted as \ac{DNC}-(\ac{DNN}) and \ac{DNC}-(\ac{GRU}), respectively. 

\subsubsection{``Classical \acsp{RNN}'': \acsp{GRU} \cite{cho2014learning} and \acsp{LSTM} \deactreview{\cite{hochreiter1997long}}}
\label{sec:GRUs}

\begin{figure}
\centering
\begin{tikzpicture}
\tikzstyle{box} = [draw,rounded corners=.1cm, align=center,fill=white] 
\tikzstyle{circ} = [draw=anthrazit,circle,minimum size=1.2cm, text centered,align=center] 

\def\width{4};

\draw[very thick,lila,rounded corners=.1cm,fill=lila!03] (-4,2.75) rectangle (4,-2.25);

\foreach \x in {0,...,8}
	\node[box,rotate=90] (1gru\x) at (\x*0.75-0.75*\width,1.5) {GRU};

\node (1fwd0) at ([yshift=4]-0.9*\width,1.5) {\small $\mathbf{0}$};
\node (1bwd0) at ([yshift=-4]0.9*\width,1.5) {\small $\mathbf{0}$};
\node (1fwdend) at ([yshift=4]0.9*\width,1.5) {};
\node (1bwdend) at ([yshift=-4]-0.9*\width,1.5) {};

\foreach \x in {0,...,8}
	\node[box,rotate=90] (2gru\x) at (\x*0.75-0.75*\width,0) {GRU};

\node (2fwd0) at ([yshift=4]-0.9*\width,0) {\small $\mathbf{0}$};
\node (2bwd0) at ([yshift=-4]0.9*\width,0) {\small $\mathbf{0}$};
\node (2fwdend) at ([yshift=4]0.9*\width,0) {};
\node (2bwdend) at ([yshift=-4]-0.9*\width,0) {};

\foreach \x in {0,...,8}
	\node[box,rotate=90] (3gru\x) at (\x*0.75-0.75*\width,-1.5) {GRU};

\node (3fwd0) at ([yshift=4]-0.9*\width,-1.5) {\small $\mathbf{0}$};
\node (3bwd0) at ([yshift=-4]0.9*\width,-1.5) {\small $\mathbf{0}$};
\node (3fwdend) at ([yshift=4]0.9*\width,-1.5) {};
\node (3bwdend) at ([yshift=-4]-0.9*\width,-1.5) {};

\foreach \x in {0,...,8}
	\draw[->,dunkelgrau] (\x*0.75-0.75*\width,3) -- (1gru\x.east);

\foreach \x [count=\xi] in {0,...,7}
	\draw[<-,rot] ([yshift=4]1gru\xi.north) -- ([yshift=4]1gru\x.south);
\foreach \x [count=\xi] in {0,...,7}
	\draw[->,mittelblau] ([yshift=-4]1gru\xi.north) -- ([yshift=-4]1gru\x.south);

\draw[->,rot] ([xshift=4]1fwd0.center) -- ([yshift=4]1gru0.north);
\draw[->,mittelblau] ([xshift=-4]1bwd0.center) -- ([yshift=-4]1gru8.south);

\foreach \x in {0,...,8}
	\draw[->,mittelblau] ([xshift=3]1gru\x.west) -- ([xshift=3]2gru\x.east);
\foreach \x in {0,...,8}
	\draw[->,rot] ([xshift=-3]1gru\x.west) -- ([xshift=-3]2gru\x.east);

\foreach \x [count=\xi] in {0,...,7}
	\draw[<-,rot] ([yshift=4]2gru\xi.north) -- ([yshift=4]2gru\x.south);
\foreach \x [count=\xi] in {0,...,7}
	\draw[->,mittelblau] ([yshift=-4]2gru\xi.north) -- ([yshift=-4]2gru\x.south);

\draw[->,rot] ([xshift=4]2fwd0.center) -- ([yshift=4]2gru0.north);
\draw[->,mittelblau] ([xshift=-4]2bwd0.center) -- ([yshift=-4]2gru8.south);

\foreach \x in {0,...,8}
	\draw[->,mittelblau] ([xshift=3]2gru\x.west) -- ([xshift=3]3gru\x.east);
\foreach \x in {0,...,8}
	\draw[->,rot] ([xshift=-3]2gru\x.west) -- ([xshift=-3]3gru\x.east);

\foreach \x [count=\xi] in {0,...,7}
	\draw[<-,rot] ([yshift=4]3gru\xi.north) -- ([yshift=4]3gru\x.south);
\foreach \x [count=\xi] in {0,...,7}
	\draw[->,mittelblau] ([yshift=-4]3gru\xi.north) -- ([yshift=-4]3gru\x.south);

\draw[->,rot] ([xshift=4]3fwd0.center) -- ([yshift=4]3gru0.north);
\draw[->,mittelblau] ([xshift=-4]3bwd0.center) -- ([yshift=-4]3gru8.south);

\foreach \x in {0,...,8}
	\draw[->,mittelblau] ([xshift=3]3gru\x.west) -- ([xshift=3]\x*0.75-0.75*\width,-2.5);
\foreach \x in {0,...,8}
	\draw[->,rot] ([xshift=-3]3gru\x.west) -- ([xshift=-3]\x*0.75-0.75*\width,-2.5);

\node[anthrazit,fill=lila!02,inner sep=2.3] at (0,2.4) {\large Multi-RNN Cell};

\end{tikzpicture}
\caption{Final Multi-\ac{RNN} cell based on \acp{GRU}.}
\label{fig:multi_rnn_cell}
\vspace*{-0.2cm}
\end{figure}
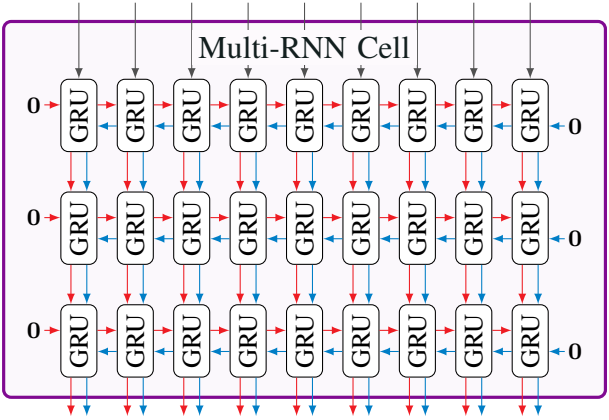

As will be shown in the results, these cell structures provide sufficient complexity to solve the task while outperforming all other recurrent structures we tested in terms of low complexity and training convergence speed.
The \ac{GRU}-based multi-\ac{RNN} cell used for all experiments throughout this work is depicted in Fig.~\ref{fig:multi_rnn_cell}.
It consists of 3 \ac{GRU} layers with $256$ units per layer and \ac{tanh} activation function.
One further advantage is the availability of CuDNN (an Nvidia library) implementations within the tensorflow library that are highly optimized for \ac{GPU} computations.
We observe a reduction of up to factor 100 in training time compared to other \ac{RNN} implementations.
This renders the possibility of more training steps within the same amount of time.
We also chose to use \acp{GRU} over \acp{LSTM} due to the implementational ease of only holding one internal state.

\section{Training Methodology}

We now provide some insights in different training strategy approaches.
All presented results of this section's experiments are based on decoding the standard non-systematic memory $\nu = 6$ convolutional code $(\text{o}133,\text{o}171)_6$. %

\subsection{Deep Learning Basics}

\subsubsection{Loss Function}

As we want the \ac{NN} to give an estimation of whether a bit was 0 or 1, we face a binary classification problem where the \ac{NN}'s output is a vector of probabilities $\hat{\pv} = \left[ P\left(\hat{u}_1 = 1\right),  P\left(\hat{u}_2 = 1\right), \dots, P\left(\hat{u}_{\ell_\text{ld}} = 1\right) \right]$.
Thus, we define the loss $J$ as binary cross-entropy (or log-loss) function
\begin{equation}
J_\text{log} = - \sum_{k=1}^{\ell_\text{ld}} u_k \cdot \log{\hat{p}_k} + \left( 1 - u_k \right) \cdot \log{ \left( 1 - \hat{p}_k \right) }
\label{eq:bce}
\end{equation}
where $\ell_\text{ld}$ is the length of the bit sequence that is contributing to the loss (loss depth).
By introducing $\ell_\text{ld} > 1$ we speed up the training procedure by reducing state ramp-up overhead that would otherwise occur by simply increasing the amount of $N$ samples within a mini-batch.
This also increases the final \ac{BPTT} depth of the gradient, which is at least of depth $\ell_\text{ramp}$ and at most of depth $\ell_\text{ramp} + \ell_\text{ld}-1$ time steps.

\deactreview{It is worth mentioning, that we also achieved similar results in terms of convergence speed by considering this problem as a regression task.
For this, the final \ac{DNN} layer is required to be linearly activated and the loss is defined as the mean squared error (L2 loss).
\begin{equation}
J_\text{mse} = \sum_{k=1}^{\ell_\text{ld}} \left( \hat{p}_k - u_k \right)^2
\label{eq:mse}
\end{equation}
}

\subsubsection{Optimizer and Learning Rate}

Both the RMSProp and Adam \deactreview{\cite{kingma2014}} optimizer were tested, with RMSProp delivering slightly better results if binary cross-entropy loss \eqref{eq:bce} is used, and Adam if the L2 loss \eqref{eq:mse} is used, respectively.
\deactreview{However, more important than the choice of the particular optimizer, is the value of the learning rate $\eta$, especially for codes with high memory.
If the $\eta$ is too high during in-depth training, we often noticed abrupt increases of training loss, sometimes leading to a complete loss of generalization.}
Throughout this work, we opt for a slow learning rate of $\eta=10^{-4}$.

\subsubsection{Metrics}

The most obvious metric, besides loss, to evaluate the performance of the \ac{NN}-based decoder is the \ac{BER}.
We calculate the \ac{BER} of a mini-batch by hard decision as in
\begin{equation}
\ac{BER} = \mathbb{E}\left[ \frac{1}{\ell_\text{ld}} \sum_{k=1}^{\ell_\text{ld}} \mathbbm{1}_{\{ \left(\hat{p}_k > 0.5 \right) \neq u_k \} } \right].
\end{equation}
As our decoder predicts $\ell_\text{ld}$ bits at once in a sample, one has to take care of the individual \ac{BER} of each bit during architecture design.
The \ac{BER} of a specific bit within a mini-batch is calculated by
\begin{equation}
\ac{BER}_k = \mathbb{E}\left[ \mathbbm{1}_{\{ \left(\hat{p}_k > 0.5 \right) \neq u_k \} } \right]
\end{equation}
and, therefore, a significant inequality of \ac{BER} between bits of different spatial positions (e.g., $\ac{BER}_0 > \ac{BER}_{\nicefrac{\ell_\text{ld}}{2}}$) is a clear indicator that the hyperparameter $\ell_\text{ramp}$ was chosen to small.

Another informative metric is the \ac{NVE} \cite{gruber2017deep}.
Since there exists an \ac{ML} decoder, we can also evaluate our \ac{NN}-based decoder's performance by normalizing its \ac{BER} within a certain \ac{SNR} range to the optimal achievable \ac{BER} obtained by the  Viterbi decoder within this \ac{SNR} range.
This metric was introduced in a similar way in \cite{gruber2017deep} and is defined as
\begin{equation}
\acs{NVE} \left( \rho \right) = \frac{1}{S} \sum_{s=1}^S \frac{\ac{BER}_\text{NND} \left( \rho , \rho_{\text{SNR},s} \right)}{\ac{BER}_\text{Viterbi} \left( \rho_{\text{SNR},s} \right)}
\end{equation}
where $\rho$ is the design parameter of the \ac{NN} that shall be investigated, $\rho_{\text{SNR}}$ denotes the \ac{SNR} and $S$ is the number of \ac{SNR} points.
The \ac{NVE} provides an easy to understand metric that depicts the influence of a certain parameter with respect to the optimal performance.

\subsection{A Priori Ramp-Up Training}

While convolutional codes up to memory $\nu = 4$ are easy to train with the presented \ac{NN}-based decoder, we struggled (or never managed) to achieve convergence for codes with higher memory.
In order to mitigate this problem we propose a pre-training method which can hopefully be adopted for many more sequence-based decoding problems, coined \emph{a priori ramp-up} training.
The basic idea is that, instead of starting the training with an equal distribution of zeros and ones in $\uv$ where $P_\text{ap} \left( u_k = 1 \right) = \nicefrac{1}{2}$, we start training with a distribution that favors either zeros or ones, i.e. $P_\text{ap} \left( u_k = 1 \right) < \nicefrac{1}{2}$.
In the context of Information Theory, this is equal to lowering the entropy of the sequence snippet $\uv$ during the beginning of the training process and then gradually increasing the entropy of $\uv$ until it reaches its maximum at $P_\text{ap} \left( u_k = 1 \right) = \nicefrac{1}{2}$.
It can also be interpreted as statistically reducing the available codeword space similar to what has been done in \cite{gruber2017deep} where a clear separation of codewords that have been used for training and inference was enforced.
This process variably reduces complexity and, thereby, makes it easier for the \ac{NN} to learn the decoding scheme in opposite to beginning training with the full codebook.
The perfect amount of \emph{a priori ramp-up} during training is still part of current research, but in the following we present three different approaches:

\begin{itemize}
\item \emph{linear} -- Gradually increase $P_\text{ap} \left( u_k = 1 \right)$ for each training step.
\item \emph{stepwise} -- Maintain a constant $P_\text{ap} \left( u_k = 1 \right)$ over several training steps and then increase it after a certain criteria is reached.
\item \emph{abrupt} -- Begin training at a certain constant level, e.g., $P_\text{ap} \left( u_k = 1 \right) = 0.1$, and then, after a certain criteria is reached, we continue training at $P_\text{ap} \left( u_k = 1 \right) = \nicefrac{1}{2}$.
\end{itemize}

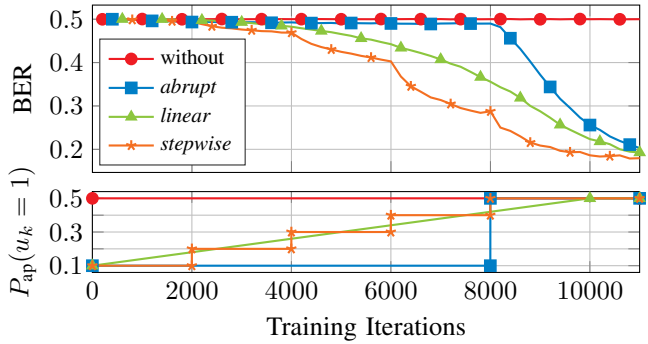
\begin{figure}
\centering
\begin{tikzpicture}
\begin{axis}[
        width=\linewidth,
	    height=0.43\linewidth,
        xmajorgrids,
        yminorticks=true,
        ymajorgrids,
        yminorgrids,
        legend style={legend cell align=left,align=left,draw=white!15!black, font=\footnotesize,at={(0.28,0.8)}},
        xticklabels={,,},
        xmin=0,
        xmax=11000,
        ylabel={BER},
        cycle list name=corporate colours markers,
		scaled x ticks = false,
		every axis plot/.append style={thick}
    ]

\addplot+[mark repeat=4,mark phase=1] table {data/ramp_up/ber_without.txt};
\addlegendentry{without};

\addplot+[mark repeat=4,mark phase=2] table {data/ramp_up/ber_abrupt.txt};
\addlegendentry{\emph{abrupt}};

\addplot+[mark repeat=4,mark phase=3] table {data/ramp_up/ber_linear.txt};
\addlegendentry{\emph{linear}};

\addplot+[mark repeat=4,mark phase=4] table {data/ramp_up/ber_stepwise.txt};
\addlegendentry{\emph{stepwise}};

\end{axis}

\begin{axis}[
        width=\linewidth,
	    height=0.3\linewidth,
        xmajorgrids,
        yminorticks=true,
        ymajorgrids,
        yminorgrids,
        xlabel={Training Iterations},
        xmin=0,
        xmax=11000,
        xtick={0,2000,4000,6000,8000,10000},
        xticklabels={$0$,$2000$,$4000$,$6000$,$8000$,$10000$},
        ytick={0.1,0.2,0.3,0.4,0.5},
        yticklabels={$0.1$,,$0.3$,,$0.5$},
        ylabel={$P_\text{ap}(u_k=1)$},
		ylabel shift = -2,
		at={(0,-0.15\linewidth)},
        cycle list name=corporate colours markers,
		scaled x ticks = false,
		every axis plot/.append style={thick}
    ]

\addplot coordinates {
(0 , 0.5)
(11000, 0.5)
};

\addplot coordinates {
(0 , 0.1)
(7999 , 0.1)
(8000 , 0.5)
(11000, 0.5)
};

\addplot coordinates {
(0 , 0.1)
(10000 , 0.5)
(11000, 0.5)
};

\addplot coordinates {
(0 , 0.1)
(1999 , 0.1)
(2000 , 0.2)
(3999 , 0.2)
(4000 , 0.3)
(5999 , 0.3)
(6000 , 0.4)
(7999 , 0.4)
(8000 , 0.5)
(11000, 0.5)
};

\end{axis}

\end{tikzpicture}
\caption{\ac{BER} performance at $E_b/N_0 = 1.5$dB over the initial training iterations for different \emph{a priori ramp-up} approaches.}
\label{fig:ramp_up}
\vspace*{-0.2cm}
\end{figure}

Fig.~\ref{fig:ramp_up} shows the \ac{BER} performance at $E_b/N_0 = 1.5$dB for different \emph{a priori ramp-up} training approaches during the initial $11,000$ training iterations.
As can be seen from the result without the use of \emph{a priori ramp-up} training, the \ac{NN}-based decoder is not able to generalize to the problem of decoding the $(\text{o}133,\text{o}171)_6$ code at all, as the \ac{BER} does not decrease but constantly stays at $0.5$ throughout the training.
This also does not change for excessively more training iterations, because the initial \emph{barrier-of-entry}, being the complexity of the full codebook, is prohibitively large. %
\emph{A priori ramp-up} is therefore needed to initialize a learning behavior at all.
This can be seen for all other approaches where \emph{a priori ramp-up} training is used.
While initially training with a low $P_\text{ap} \left( u_k = 1 \right)$ and then \emph{abruptly} increasing to $P_\text{ap} \left( u_k = 1 \right) = \nicefrac{1}{2}$ is already sufficient to spark a convergence for further training at $P_\text{ap} \left( u_k = 1 \right) = \nicefrac{1}{2}$, we can see that the \emph{linear} and \emph{stepwise} approaches further increase the learning speed in terms of fewer iterations needed to achieve the same \ac{BER} performance.
This is why we use \emph{stepwise} \emph{a priori ramp-up} training throughout this work.

\subsection{Important Hyperparameters}

\subsubsection{Layer Dimensions}

\deactreview{In general, parameterization concerning the amount of recurrent and dense layers and their respective amount of units is heavily dependent on the convolutional code.}
The \ac{RNN} cell parameters we mention in Section~\ref{sec:GRUs} were used to decode the $(\text{o}133,\text{o}171)_6$ convolutional code.
\deactreview{For less complex codes with fewer memory, less layers and less units are equally sufficient.
Also, different amounts of units per \ac{RNN} layer are possible, but to be able to use highly performance optimized CuDNN \ac{GRU} layer implementations in tensorflow, equal amounts of units per layer are required.}
For the combining \ac{DNN} layer we use $N_\text{DNN}=16$ units.
We found out that one combining layer is enough to do the job, more \ac{DNN} layers resulted in a slower convergence during training throughout our experiments.

\subsubsection{Training \ac{SNR}}

In \cite{gruber2017deep} it has been empirically shown that an optimal training SNR exists for a given code and \ac{NN} architecture.
This has also been shown analytically later in \cite{Benammar2018optimal}.
The intuition behind is that the optimal training SNR is a trade-off between training only the code structure (i.e., the inverse encoding function in the noiseless case) and learning how to handle noisy observations.
We found, empirically, that the optimal \ac{SNR} during training is at the point where the convolutional code performs in a range of $\ac{BER} = 10^{-1}$ to $\ac{BER} = 10^{-2}$.
For most of the codes we investigated, this means a training \ac{SNR} range between $1$dB to $1.5$dB.

\subsubsection{Traceback Length}

As mentioned before, the design parameters $\ell_\text{ramp}$ and $\ell_\text{ld}$ are highly important and must be matched to the convolutional code.
While $\ell_\text{ld}$ is only used to improve training efficiency, $\ell_\text{ramp}$ basically defines the depth of the \ac{NN}'s gradient and can thereby be interpreted as the \ac{NN}-based decoder's ``traceback'' length \cite{hemmati1977truncation}.
To be able to achieve close to \ac{ML} performance, it is important that the gradient for the predictions $p_1$ and $p_\text{ld}$ is at least $\ell_{\text{tb}}$ time steps deep in both directions.
We ensure this by setting the state ramp-up length to $\ell_{\text{ramp}} = \ell_{\text{tb}}$ for most of our experiments.

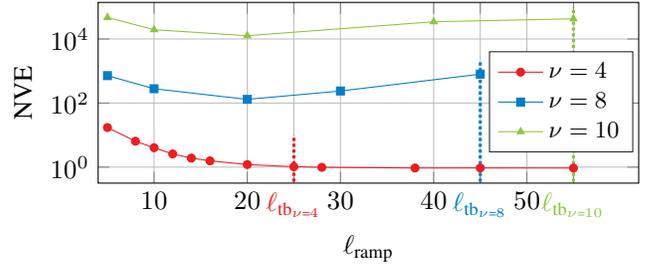
\begin{figure}
\centering
\begin{tikzpicture}
\begin{axis}[
        width=.99\linewidth,
	    height=0.45\linewidth,
        xmajorgrids,
        yminorticks=true,
        ymajorgrids,
        yminorgrids,
        legend style={legend cell align=left,align=left,draw=white!15!black, font=\small,at={(0.99,0.45)},anchor=east},
        xlabel={$\ell_{\text{ramp}}$},
        ylabel={NVE},
		ymode=log,
        xtick={10,20,25,30,40,45,50,55},
        xticklabels={$10$,$20$,\color{rot}{$\ell_{\text{tb}_{\nu\text{=}4}}$},$30$,$40$,\color{mittelblau}{$\ell_{\text{tb}_{\nu\text{=}8}}$},$50$,\color{apfelgruen}{$\ell_{\text{tb}_{\nu\text{=}10}}$}},
		xmin = 4,
		xmax = 62,
		ymin = 0,
        mark size=1.5pt,
        cycle list name=corporate colours markers
    ]	

\addplot table {data/traceback_nve/nve_m4.txt};
\addlegendentry{$\nu = 4$}
\draw[rot,very thick,densely dotted] (25,0.1) -- (25,10);

\addplot table {data/traceback_nve/nve_m8.txt};
\addlegendentry{$\nu = 8$}
\draw[mittelblau,very thick,densely dotted] (45,0.1) -- (45,2000);

\addplot table {data/traceback_nve/nve_m10.txt};
\addlegendentry{$\nu = 10$}
\draw[apfelgruen,very thick,densely dotted] (55,0.1) -- (55,99000);

\end{axis}

\end{tikzpicture}
\caption{Impact of \emph{gradient depth} on the \ac{NVE} performance over different ramp-up lengths $\ell_{\text{ramp}}$.}
\label{fig:nve_codes}
\vspace*{-0.2cm}
\end{figure}

Fig.~\ref{fig:nve_codes} depicts the impact of gradient depth by showing the \ac{NVE} over different ``traceback'' lengths $\ell_{\text{ramp}}$ while $\ell_{\text{ld}} = 1$ for this experiment (to guarantee a constant gradient depth).
To calculate a suitable \ac{NVE} we chose $S=8$ \ac{SNR}-points equally spaced starting from $\rho_{\text{SNR},1}=0$dB up to $\rho_{\text{SNR},8}=3.5$dB.
As can be seen for both codes, the decoding performance heavily depends on the gradient depth and results in a bathtub curve for too complex codes with high memory $\nu$ if plotted over different $\ell_{\text{ramp}}$.
It is obvious that while $\ell_{\text{ramp}} < \ell_{\text{tb}}$, the decoder can not reach \ac{ML} performance, but if $\ell_{\text{ramp}}$ is chosen too high, more training would be required. %
While $\ell_{\text{ramp}} \geq \ell_{\text{tb}}$ can easily be trained when decoding the $\nu=4$ code, this is not possible for the $\nu=8$ and $\nu=10$ codes since the \ac{NN} becomes too deep and complex to achieve optimal performance.
Also note that we stopped training for the $\nu=8$ and $\nu=10$ codes after several days since there was no further improvement.

\section{Results}

In this Section, we will present some results which demonstrate that it is possible to process high-entropy signals with \acp{NN}, even for highly complex tasks like decoding a memory $\nu = 6$ convolutional code as used in 802.11 \cite{ieee_wifi_n} and many other communication standards.

\subsection{Comparison Of Different \ac{RNN} Cells}

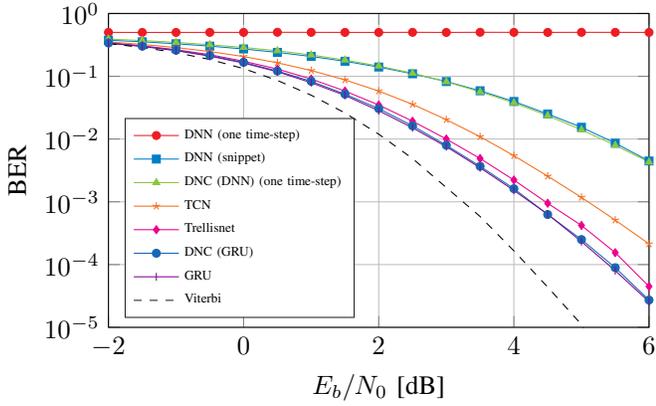
\begin{figure}
\centering
\begin{tikzpicture}
\begin{axis}[
        width=.99\linewidth,
	    height=0.65\linewidth,
        xmajorgrids,
        yminorticks=false,
        ymajorgrids,
        legend pos=south west,        
        legend style={legend cell align=left,align=left,draw=white!15!black, font=\tiny},
        xlabel={$E_b/N_0$ [dB]},
        ylabel={BER},
        ymode=log,
		xmin = -2,
		xmax = 6,
		ymin = 1e-5,
		ymax = 1,
        mark size=1.5pt,
        cycle list name=corporate colours markers
    ]	

\addplot table {data/cell_comparison/ber_paper_dnn_timestep.txt};
\addlegendentry{DNN (one time-step)}

\addplot table {data/cell_comparison/ber_paper_ff.txt};
\addlegendentry{DNN (snippet)}

\addplot table {data/cell_comparison/ber_paper_dnc_ff.txt};
\addlegendentry{DNC (DNN) (one time-step)}

\addplot table {data/cell_comparison/ber_paper_tcn.txt};
\addlegendentry{TCN}

\addplot table {data/cell_comparison/ber_paper_trellisnet.txt};
\addlegendentry{Trellisnet}

\addplot table {data/cell_comparison/ber_paper_dnc_rnn.txt};
\addlegendentry{DNC (GRU)}

\addplot table {data/cell_comparison/ber_paper_gru.txt};
\addlegendentry{GRU}

\addplot[dashed] table {data/cell_comparison/viterbi_baseline.txt};
\addlegendentry{Viterbi}

\end{axis}

\end{tikzpicture}
\caption{\ac{BER} performance of different \ac{RNN} cell structures after limited training with the $\left(\text{o}23,\text{o}35 \right)_4$ convolutional code.}
\label{fig:cell_comparison}
\vspace*{-0.2cm}
\end{figure}

Fig.~\ref{fig:cell_comparison} provides a \ac{BER} over \ac{SNR} performance comparison of all tested \ac{RNN} cell structures.
The amount of training steps is fixed to $10,000$ iterations (which is not enough to reach Viterbi performance) and all training hyperparameters are consistent throughout all structures.
We chose to present this result for the $\left(\text{o}23,\text{o}35 \right)_4$ convolutional code as some structures did not converge for higher memory codes and to limit training complexity.

\begin{enumerate}
\item As can be seen the \ac{GRU}-based multi-\ac{RNN} cell performs best after training with this limited amount of iterations.

\item Also the \ac{DNC}-(\ac{GRU}) cell, which uses \ac{GRU}-based controllers yields the same performance.
However, we assume this may be mainly caused by the embedded \ac{GRU} structure.

\item The \ac{DNC}-(\ac{DNN}) cell, using \ac{DNN}-based controllers, performs as bad as the snippet-based \ac{DNN} decoder. But this shows (and is worth mentioning), that the \ac{DNC}-(\ac{DNN}) must use its attached memory as we only feed one time-step $\yv_k$ per decision.
In contrary to the non-snippet based \ac{DNN} (i.e., a \emph{\ac{DNN} (one time-step)} that also only sees $\yv_k$ per decision) this net structure is in principle able to decode and, thus, must make use of its external memory.

\item The Trellisnet and \ac{TCN} cells also show a better convergence than the simple snippet-based forward fed \ac{DNN}, which means their structure also improves signal processing for this problem.

\end{enumerate}

From a rather practical perspective one of the most important problems with complex structures like the \ac{DNC}-(\ac{GRU}), Trellisnet and \ac{TCN} cells is, that their computation time is way longer than the \ac{GRU} cell's.
This means one can perform way more training iterations with a \ac{GRU}-cell-based decoder than with the more complex cells in the same amount of time.

\subsection{Achieved \acp{BER} for different convolutional codes}

\begin{figure}
\centering
\begin{tikzpicture}
\begin{axis}[
        width=.99\linewidth,
	    height=0.65\linewidth,
        xmajorgrids,
        yminorticks=false,
        ymajorgrids,
        legend pos=south west,        
        legend style={legend cell align=left,align=left,draw=white!15!black, font=\tiny},
        xlabel={$E_b/N_0$ [dB]},
        ylabel={BER},
        ymode=log,
		xmin = 0,
		xmax = 6,
		ymin = 1e-5,
		ymax = 1,
        mark size=1.5pt,
        cycle list name=corporate colours markers
    ]	

\addplot table {data/code_comparison/ber_nn_m1.txt};
\addlegendentry{$\nu = 1$}

\addplot table {data/code_comparison/ber_nn_m2.txt};
\addlegendentry{$\nu = 2$}

\addplot table {data/code_comparison/ber_nn_m4.txt};
\addlegendentry{$\nu = 4$}

\addplot table {data/code_comparison/ber_nn_m6.txt};
\addlegendentry{$\nu = 6$}

\addplot table {data/code_comparison/daniel_ber_long_train_8_tb_20.txt};
\addlegendentry{$\nu = 8$}

\addplot table {data/code_comparison/daniel_ber_long_train_10_tb_20.txt};
\addlegendentry{$\nu = 10$}

\addplot[dashed, no markers,rot] table {data/code_comparison/ber_stb_vit_m1.txt};

\addplot[dashed, no markers,mittelblau] table {data/code_comparison/ber_stb_vit_m2.txt};

\addplot[dashed, no markers,apfelgruen] table {data/code_comparison/ber_stb_vit_m4.txt};

\addplot[dashed, no markers,orange] table {data/code_comparison/ber_stb_vit_m6.txt};

\addplot[dashed, no markers,pink] table {data/code_comparison/ber_stb_vit_m8.txt};

\addplot[dashed, no markers,lila] table {data/code_comparison/ber_stb_vit_m10.txt};

\end{axis}

\end{tikzpicture}
\caption{\ac{BER} performance of the \ac{NN}-based decoder for different codes (Viterbi performance in dashed lines).}
\label{fig:ber_codes}

\end{figure}
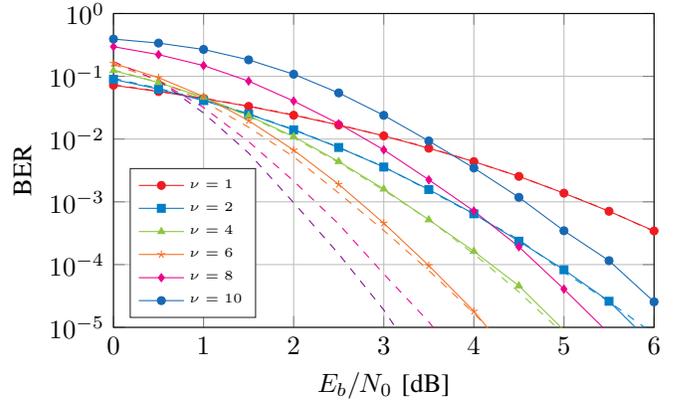

\begin{table}
	\caption{Characteristics of all learned convolutional codes}
	\centering
	\begin{tabular}{ccccc} 
	\toprule
 	type & generator polynomials & code rate & constr. length ($\nu+1$) & $\ell_\text{tb}$\\
 	\midrule 
	\={N}\={S}C & $\left(\text{o}1,\text{o}3 \right)_1$ & $\nicefrac{1}{2}$ & 2 & 10 \\
	\={N}\={S}C & $\left(\text{o}5,\text{o}7 \right)_2$ & $\nicefrac{1}{2}$ & 3 & 15 \\
	\={N}\={S}C & $\left(\text{o}23,\text{o}35 \right)_4$ & $\nicefrac{1}{2}$ & 5 & 25 \\
	\={N}\={S}C & $\left(\text{o}133,\text{o}171 \right)_6$ & $\nicefrac{1}{2}$ & 7 & 35 \\
	\={N}\={S}C & $\left(\text{o}561,\text{o}753 \right)_8$ & $\nicefrac{1}{2}$ & 9 & 45 \\
	\={N}\={S}C & $\left(\text{o}2335,\text{o}3661 \right)_{10}$ & $\nicefrac{1}{2}$ & 11 & 55 \\
	\bottomrule
	\end{tabular}
	\label{tab:codes}
	\vspace*{-0.2cm}
\end{table}

Fig.~\ref{fig:ber_codes} shows the \ac{BER} performance of the proposed \ac{NN} based decoder for different convolutional codes.
The characteristics of all investigated codes are listed in Table~\ref{tab:codes}.
It can be seen that the \ac{NN}-based decoder is able to pretty much achieve the Viterbi performance for all codes up to memory $\nu = 6$, although the performance for codes with higher memory shows a significant gap of several dB to the optimal performance.
Both results for the $\nu=8$ and $\nu=10$ codes are achieved by reducing the \ac{NN}-based decoder's ``traceback'' and, thereby, its complexity to $\ell_{\text{ramp}} = 20$.
We also stopped the training process for these codes after about two days of computing time, yet Viterbi performance may possibly be reached after even more training or parallelization approaches.
We still find it quite remarkable that, due to \emph{a priori ramp-up} training, it is possible to initiate some generalization even for extremely complex convolutional codes of memory $\nu > 6$.

\section{Joint detection and decoding}

In a \ac{BICM} scheme with non-Gray labeling, iterations between demapper and decoder are usually required to recover the full information \cite{tenbrink1998demapping}. Thus, for anti-Gray labeling and a single demapper iteration the \ac{BER} is inevitably degraded. Besides additional complexity, such iterative receiver schemes also increase the overall decoding latency significantly.
In this final Section, we show that an \ac{NN}-based decoder is inherently able to achieve a \ac{BER} performance that could otherwise only be reached using iterative demapping and decoding.
For this, we extend our system model by anti-Gray \ac{QPSK} mapping. %

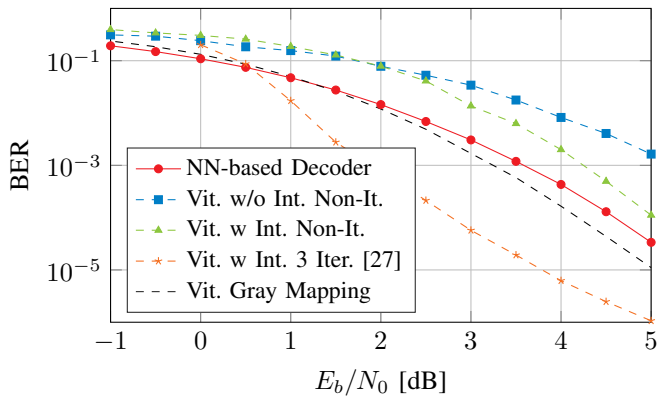
\begin{figure}
\centering
\begin{tikzpicture}
\begin{axis}[
        width=.99\linewidth,
	    height=0.65\linewidth,
        xmajorgrids,
        yminorticks=true,
        ymajorgrids,
        legend pos=south west,        
        legend style={legend cell align=left,align=left,draw=white!15!black, font=\small},
        xlabel={$E_b/N_0$ [dB]},
        ylabel={BER},
        ymode=log,
		xmin = -1,
		xmax = 5,
		ymin = 1e-6,
		ymax = 1,
        mark size=1.5pt,
        cycle list name=corporate colours markers
    ]	

\addplot table {data/antigray/ber_nn.txt};
\addlegendentry{NN-based Decoder}

\addplot+[dashed,mark options={solid}] table {data/antigray/ber_viterbi_antigray_woInt.txt};
\addlegendentry{Vit. w/o Int. Non-It.}

\addplot+[dashed,mark options={solid}] table {data/antigray/ber_viterbi_antigray_wInt.txt};
\addlegendentry{Vit. w Int. Non-It.}

\addplot+[dashed,mark options={solid}] table {data/antigray/ber_viterbi_antigray_wInt_iterative_abgelesen.txt};
\addlegendentry{Vit. w Int. 3 Iter. \cite{tenbrink1998demapping}}

\addplot[dashed,no markers] table {data/antigray/ber_viterbi_gray.txt};
\addlegendentry{Vit.  Gray Mapping}

\end{axis}

\end{tikzpicture}
\vspace*{-0.3cm}
\caption{\ac{BER} performance of the \ac{NN}-based decoder for anti-Gray mapped \ac{QPSK} modulation.}
\label{fig:ber_antigray}
\vspace*{-0.2cm}
\end{figure}

Fig.~\ref{fig:ber_antigray} shows the \ac{NN}-based decoders performance if anti-Gray labeled \ac{QPSK} modulation is used.
As can be seen it outperforms the non-iterative Viterbi decoder for both with and without using a bit-interleaver and even slightly outperforms the Gray labeling \ac{QPSK} Viterbi performance at low \ac{SNR}.
As reported in \cite{tenbrink1998demapping}, the Viterbi decoder using a bit-interleaver and iterative demapping and decoding with only 3 iterations then again easily performs better than the \ac{NN}-based decoder\deactreview{\footnote{Curve taken from \cite{tenbrink1998demapping}}}.
However, we want to emphasize that the \ac{NN}-based decoder provides \emph{one-shot} estimates outperforming the optimal non-iterative scheme. As in \cite{gruber2017deep}, we coin the term \emph{one-shot} decoding as this scheme does not need any further iterations and, thus, can possibly operate at much lower overall decoding latency.

\deactreview{
\section{Conclusion and Outlook}
In this paper, we compared several \ac{NN} architectures for sequence-based processing on the task of decoding convolutional codes.
We showed that, although all tested \acp{NN} were able to converge, the already well investigated and highly performance optimized \ac{GRU} and \ac{LSTM} cells are most suitable to tackle such exemplary complex communications related signals.
We have demonstrated that an \ac{NN}-based decoder is able to optimally decode convolutional codes up to memory $\nu \leq 6$ and introduced an original \emph{ramp up} training, which actually enables convergence for memory $\nu \geq 6$ in the first place.
Further, we showed that the proposed \ac{NN}-based decoder is able to learn what otherwise would only be possible with iterative processing schemes, which, again, underlines the high potential of \ac{NN}-based components.
}

\bibliographystyle{IEEEtran}
\bibliography{IEEEabrv,references}

\end{document}